\documentclass[twocolumn,reprint,preprintnumbers,
amsmath,amssymb,aps, showkeys]{revtex4-1}

\usepackage{graphicx}% Include figure files
\usepackage{dcolumn}% Align table columns on decimal point
\usepackage{bm}% bold math
\usepackage{amssymb}
\usepackage{amsmath}
\usepackage{color}
\usepackage{soul}
 \usepackage{float}
\usepackage{subfig}
\usepackage{hyperref}% add hypertext capabilities

\begin{document}
\title{Phase Transitions in the Anisotropic Dicke-Stark Model with $\mathbf{A}$-square
terms}
\author{Xiang-You Chen$^{1}$, Yu-Yu Zhang$^{2}$, Qing-Hu Chen$^{3,*}$, and Hai-Qing Lin$^{1,3,4,\dagger}$}
\affiliation{$^{1}$ Beijing Computational Science Research Center, Beijing 100193,
China ~\\
 $^{2}$ Department of Physics, Chongqing University, Chongqing 401330,
China~\\
 $^{3}$ School of Physics, Zhejiang University, Hangzhou 310027,
China\\
$^{4}$Institute for Advanced Study in Physics of Zhejiang University,
Hangzhou, 310058, China
}

\date{\today }

\begin{abstract}
The superradiant phase transition (SRPT) is forbidden in the standard
isotropic Dicke model due to the so-called no-go theorem induced by $\mathbf{A}$-square
term. In the framework of the Dicke model, we demonstrate that SRPTs can
occur at both zero and finite temperatures if we intrinsically tune the
rotating wave and count-rotating atom-cavity coupling independently, and/or
introduce the nonlinear Stark coupling terms, thus overcoming the no-go
theorem. The phase transitions in this so-called anisotropic Dicke-Stark
model share the same universality class with the original Dicke model. The critical coupling strength of this model decreases with the anisotropic parameter  gradually, but can be driven to zero quickly with the strong
nonlinear Stark coupling. We believe that we have proposed a feasible scheme
to observe the SRPT in the future solid-state experiments.
\end{abstract}

\pacs{42.50.Pq, 68.35.Rh, 05.30.Rt, 64.75.-g}
\maketitle

\section{ Introduction}

The Dicke model describes a collection of $N$ two-level atoms interacting
with a single radiation mode via an atom-field coupling \cite{DM}. The
thermal phase transition (TPT) from the normal phase to the superradiant
phase in equilibrium was predicted more than 50 years ago \cite{DM_PT,DM_PT2}%
. In 2003, the quantum phase transition (QPT) was also theoretically studied 
\cite{emary2003prl}, which exhibit the mean-field nature. However both TPT
and QPT in equilibrium have never been convincingly observed in the
experiments which are solely described by the original Dicke model. Although
it was reported that the superradiant phase transition (SPT) was observed in
Bose-Einstein condensates (BEC) in optical cavities \cite{Baumann} and
simulation via cavity-assisted Raman transitions \cite{Zhiqiang}, it was
shown later that these two realizations of the superradiant transition in
the Dicke model involve driven-dissipative systems, thus they are not the
equilibrium but the non-equilibrium phase transitions \cite{Torre2018intro}.
By analyzing the average number of photons, it was found in Ref. \cite{Dalla}
that the critical scaling exponent for photons in the driven-dissipative
Dicke model is different from that in equilibrium Dicke model \cite{Liu}. It
follows that the emergence of the photons in experiments is not a sufficient
evidence of the equilibrium phase transitions.

One of the major challenge to realize the equilibrium superradiant phase
transition (SRPT) is that it only occurs at the very strong atom-cavity
coupling, which seems almost not accessible in the recent stongly
light-matter interaction systems based on some advanced solid-state
platforms, except in the quantum simulations. To takle this problem, two of
the present authors and collaborators have introduced the interaction
between atoms \cite{taoliu2012pla} to reduce the critical strength.

The most serious difficulty is however rooted in a fundamental issue that
the dipole coupling between the atoms and the cavity may be incompletely
considered in the standard Dicke model. So even from a theoretical
perspective, it is still in debate to realize the SRPT experimentally \cite%
{A2original,keeling2007jpb,vukics2014prl,ciuti2010nc,viehmann2011prl,ciuti2012prl,viehmann_arxiv,rabl2016pra,bamba2016prl,bamba2017pra,andolina2019prb,nori2019np,nazir2020prl,nazir2022rmp}%
. In the cavity quantum electrodynamical (QED), Rza\.{z}eskimo et al. argued
that the so-called $\mathbf{A}$-square term originated from minimal coupling
Hamiltonian would forbid the SRPT of the Dicke model at any finite coupling
strength if the Thomas-Reich-Kahn (TRK) sum rule for atom is taken into
account properly~\cite{A2original}. According to the effective model of the
circuit QED, Nataf and Ciuti proposed that the TRK sum rule of the cavity
QED could be violated and the SRPT may occur in principle \cite{ciuti2010nc}%
. Viehmann et al. questioned that even within a complete microscopic
treatment, the no-go theorem of the cavity QED still applies to the circuit
QED \cite{viehmann2011prl}. These arguments focus on whether the TRK sum
rule changes in different systems \cite{viehmann2011prl,ciuti2012prl}, which
arouse a huge controversy whether the $\mathbf{A}$-square term can be engineered in
the superconductive qubit circuit platform \cite%
{viehmann_arxiv,rabl2016pra,bamba2016prl,andolina2019prb}. A specific
circuit configuration was also designed to generate the SRPT \cite%
{bamba2016prl,bamba2017pra}. On the other hand, because the validity of the
two-level approximation is the key step of constructing the Dicke model in
various solid-state devices, the applicability of Coulomb and electric
dipole gauges, and different experimental schemes leading to different $A$%
-square terms are also widely discussed in the literature ( see \cite%
{nori2019np,Stokes,Andolina}, and the references therein). Consensus has not
been reached yet, and the no-go theorem is still the most controversial
subject.

We would not join the debate of the existence of the $\mathbf{A}$-square terms
subject to the TRK sum rule. In the present work, keeping such a $\mathbf{A}$-square
term, we propose an effective scheme to realize the SRPT in the original
Dicke model at weak atom-cavity coupling by manipulating the inner factors.
In the framework of the standard Dicke model which only consists of two
ingredients: two-level atoms without mutual interactions and single-mode
cavity, what we can do is to modify the atom-cavity interactions. In doing
so, we first relax the isotropic atom-cavity interaction to the anisotropic
one, i.e. the coupling strengths of the rotating and counterrotating wave
terms are different.Such an anisotropic Dicke model can be implemented in
the cavity and circuit QED \cite{Ciute2,mxliu2017prl}. On the other hand,
Grimsmo and Parkins proposed that a nonlinear atom-cavity coupling, nowadays
called the Stark coupling, can be realized in an experimental set-up with
two hyperfine ground states of a multilevel atom coupled to a quantized
cavity field and two auxiliary laser fields. Generalizations of the Stark
coupling into the Dicke model gives the so-called Dicke-Stark model (DSM).
It is found in \cite{KeelingPRL2010,KelingPRA2012} that the critical
coupling strength of the none-equilibrium QPT of the DSM is very sensitive
to the nonlinear Stark interaction between atoms and cavity. So we then
incorporate the Stark coupling terms in the anisotropic Dicke model and try
to reduce the critical strength of the equilibrium SRPT, so that the SRPT
may be feasible in the future advanced experiments.

The paper is organized as follows: In Sec. II, we introduce the Hamiltonian
for anisotropic Dicke-Stark model with $A$-square terms . In Sec. III, within the Holstein-Primakoff transformation, we will derive the mean-field
critical coupling strength. We will derive a free energy of the general model and discuss the thermal phase transitions in Sec. IV. Finally,
conclusions are drawn in Sec. V.

\section{The anisotropic Dicke-Stark model with $\mathbf{A}$-square term}

We consider an anisotropic Dicke-Stark model, which describes a collection
of $N$ two-level systems coupled to a bosonic field by adding a stark shift
term to the anisotropic Dicke model. The Hamiltonian is given by 
\begin{align}
H & =H_{AD}+\frac{U}{N}a^{\dagger}aJ_{z}+D(a^{\dagger}+a)^{2}, \\
H_{AD} & =\omega a^{\dagger}a+\Delta J_{z}+\frac{g}{\sqrt{N}}%
(a^{\dagger}J_{-}+aJ_{+})  \notag \\
& +\frac{g\tau}{\sqrt{N}}(a^{\dagger}J_{+}+aJ_{-})  \label{H2}
\end{align}
where $H_{AD}$ is the anisotropic Dicke Hamiltonian, and $U$-term is the
stark shift term for the nonlinear atom-cavity interactions. $a^{\dagger}(a)$
creates (annihilate) one photon in the common single-mode cavity with
frequency $\omega$, $\Delta$ is the transition frequency of each two-level
atom. The angular momentum operator represents the collective pesudospin
associated to the collection of $N$ two-level atoms, $J_{z}=\frac{1}{2}%
\sum_{i=1}^{N}\sigma_{z}^{(i)},J_{\pm}=\frac{1}{2}\sum_{i=1}^{N}\sigma_{%
\pm}^{(i)}$. The counter-roatating-wave (CRW) interacting terms $%
a^{\dagger}J_{+}$ and $aJ_{-}$ do not conserve the number of excitations,
while the rotating-wave (RW) terms $a^{\dagger}J_{-}$ and $aJ_{+}$ lead to
conserved excitations. The anisotropic coupling is tuned by $\tau$, which is
the ratio of the CRW to the RWA coupling strength $g$, which plays a crucial
role in the SRPT.

Note that the $D$ term of the Hamiltonian represents an $\mathbf{A}$-square
term imposed by the TRK sum rule for the atoms, which originates from $(%
\mathbf{p}-e\mathbf{A})^{2}/2m$ term for an atom coupled to an
electromgnetic field with a vector potential $\mathbf{A}$. For convenience
we set $D=\kappa g^{2}/\Delta$ with a dimensionless parameter $\kappa$, for
which it is required $\kappa\geqslant1$ to satisfy the TRK sum rule~\cite%
{A2original}. The $\mathbf{A}$-square term, often overlooked in other works,
will become crucial for the SRPT in the DSM with the additional stark shift
term.

%\begin{align}
%H & =\omega a^{\dagger}a+\frac{U}{N}\sum_{j=1}^{N}\sigma_{j}^{z}a^{%
%\dagger}a+\Delta\sum_{j=1}^{N}\sigma_{j}^{z}+\frac{g_{1}}{\sqrt{N}}%
%(a^{\dagger}\sigma_{j}^{-}+a\sigma_{j}^{-})  \notag \\
%& +\frac{g_{2}}{\sqrt{N}}(a^{\dagger}\sigma_{j}^{+}+a\sigma_{j}^{-})+D(a^{%
%\dagger}+a)^{2}  \label{HaniDSA2}
%\end{align}

\section{Quantum phase transitions}

In order to explore the phase diagram, we perform the Holstein-Primakoff
transformation. The angular momentum operators are expressed in terms of
bosoic operators $b$ and $b^{\dagger}$, giving $J_{+}=b^{\dagger}\sqrt{%
N-b^{\dagger}b}$, $J_{-}=\sqrt{N-b^{\dagger}b}b$, an $J_{z}=b^{\dagger}b-%
\frac{N}{2}$. Then we shift the bosonic operators with respect to their mean
value as $c^{\dagger}=a^{\dagger}-\sqrt{N}\alpha$ and $d^{\dagger}=b^{%
\dagger}-\sqrt{N}\varsigma$. The nonzero value of $\alpha$ and $\varsigma$
are the order parameters of the SRPT, which signal a macroscopic excitation
of photons and a spontaneous pseudospn polarization of the two-level atoms.
The ground-state energy are obtained by keeping the terms proportional to $N$
\begin{align}
\frac{E(\alpha,\varsigma)}{N} & =(\omega-\frac{U}{2})\alpha^{2}+U\alpha^{2}%
\varsigma^{2}  \notag \\
& +\Delta(\varsigma^{2}-\frac{N}{2})+2g(1+\tau)\alpha\varsigma+4\kappa\frac{%
g^{2}}{\Delta}\alpha^{2}
\end{align}
Minimizing the ground state energy with respect to $\alpha$ and $\varsigma$
leads to 
\begin{equation}
\alpha=\frac{2\varsigma\sqrt{1-\varsigma^{2}}\Delta g(\tau+1)}{%
2\Delta\omega+8g^{2}\kappa+\left(2\varsigma^{2}-1\right)U\Delta}
\label{alpha}
\end{equation}
and 
\begin{equation}
\varsigma=\sqrt{\frac{1}{2}-\frac{\Delta\omega+4g^{2}\kappa}{U\Delta}-\mu}
\end{equation}
where $\mu=\frac{\sqrt{g^{^{\prime }2}[g^{^{\prime
}2}+U\Delta][4\left(\Delta\omega+4g^{2}\kappa\right)^{2}-U^{2}\Delta^{2}]}}{%
2U\Delta[g^{^{\prime }2}+U\Delta]}$ with $g^{\prime }=g(\tau+1)$. The
nontrivial solutions of $\alpha$ and $\varsigma$ are obtained in the
superradiant phase, while they are zero in the normal phase. We see that the
mean value $\varsigma$ is real only when $\frac{1}{2}-\frac{%
\Delta\omega+4g^{2}\kappa}{\Delta U}-\mu\geq0$. It yields the critical
coupling strength of the SRPT 
\begin{equation}
g_{c0}^{A_{2}}=\frac{\sqrt{\Delta\omega-\frac{\Delta U}{2}}}{\sqrt{%
(\tau+1)^{2}-4\kappa}}.  \label{gc_ADS}
\end{equation}

Note that for the isotropic Dicke model $\ \tau=1$, under the TRK sum rule ($%
\kappa>1$), no phase transition can occur at a finite coupling strength,
which is consistent with the no-go theorem. Very interestingly, for the
anisotropic Dicke model, we do find the quantum phase transitions at a
finite critical coupling strength under the condition $(\tau+1)^{2}-4%
\kappa>0,i.e.,\tau>2\sqrt{\kappa}-1>1$, even constrained by the TRK sum
rule. Moreover, the critical coupling strength $g_{c0}^{A_{2}}$ decreases
with $U/\omega$, which requires $U/\omega<2$. It indicates that the presence
of the nonlinear Stark type atom-cavity interaction can reduce the critical
point, demonstrating the feasible superradiant QPT in experiments at the
weak atom-cavity coupling.

\begin{figure}[tbph]
\centerline{\includegraphics[scale=0.35]{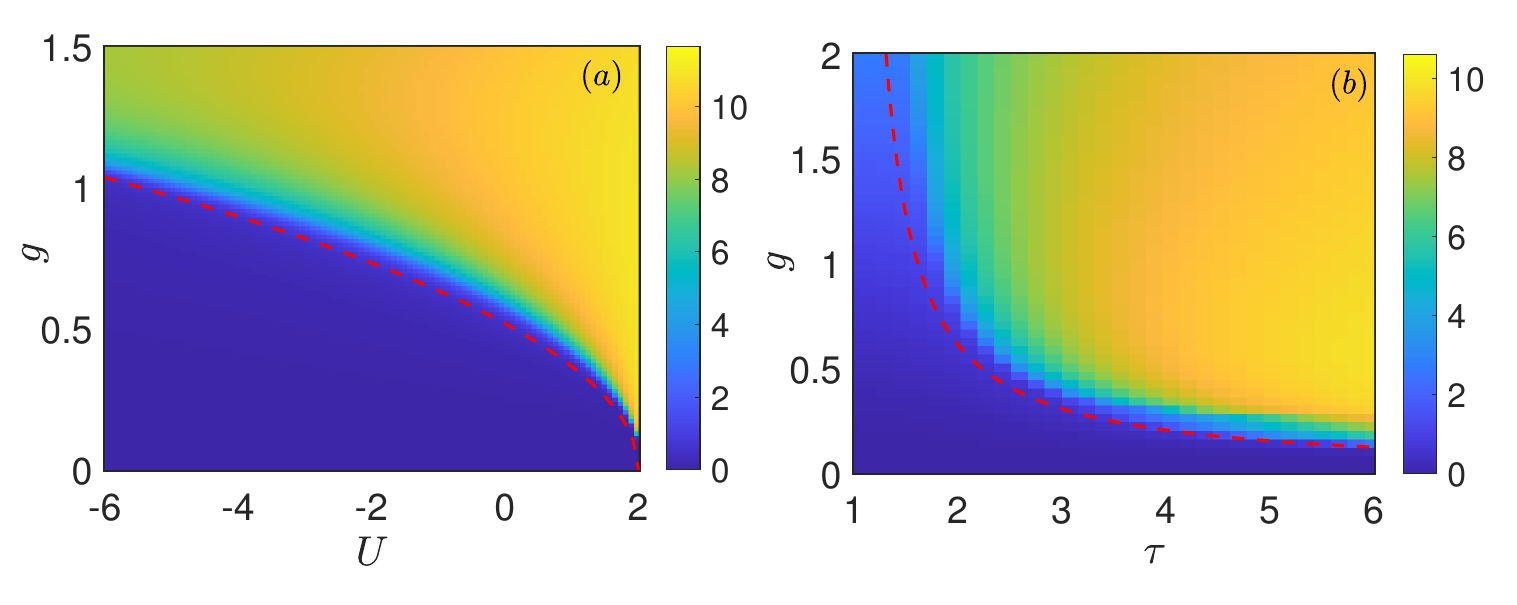}}
\caption{ The phase diagram in the $g-U$ plane for $\Delta/\protect\omega=2$%
, $\protect\tau=2.5,$ and $\protect\kappa=1.2$ (a) and in the $g-\protect%
\tau $ plane for $\Delta/\protect\omega=1/2$, $U=0.5,$ and $\protect\kappa%
=2.5$ (b) by calculating the average photon number $N_{ph}$ numerically with 
$N=200 $ atoms. The phase boundary is shown in the red dashed line by the
analytical expression in Eq.~(\protect\ref{gc_ADS}).}
\label{gc0}
\end{figure}

Fig.~\ref{gc0} shows phase diagrams in the $g-U$ and $\ g-\tau $ planes of
the DSM, respectively. We calculate the mean photon number $%
N_{ph}=\left\langle a^{\dagger }a\right\rangle $ by numerically exact
diagonalizations with $N=200$ atoms. $N_{ph}$ is almost zero below the
critical line, and then smoothly increases in the superradiant phase regime.
The analytical coupling strength $g_{c0}^{A_{2}}$ dependent on $U$ in Eq.~(\ref{gc_ADS}) (red dashed line) fits well with the critical phase boundary
in Fig.~\ref{gc0} (a). $g_{c0}^{A_{2}}$ decreases quickly when $U$
approaches to $2$. Although the nonlinear Stark coupling is not necessary
for the occurrence of the SRPT, it can effectively reduce the critical
coupling point in the practical experiments. Moreover, the critical coupling
strength $g_{c0}^{A_{2}}$ increases sharply as the anisptropic ratio tends
to $\tau =1$ in Fig.~\ref{gc0} (b). As an evidence the SRPT is absence for
the isotrpic Dicke model due to the restriction of the TRK rule. However, it
is possible to observe the emergence of the SRPT by enhancing the
anisptropy. The critical line agrees well with the analytical $g_{c0}^{A_{2}}
$. Consequence the SRPT is accessible with the additional Stark coupling for
the anisotropic Dicke model.

It is worth noting that, for $U<-2$, there still exists the superradiant
QPT, in sharp contrast to the single-atom Rabi-Stark model where the system
is in the continuum in this regime \cite{Rabi_stark}. It may underline the
many-body effect in the Dicke-Stark model. For $U>2$, the numerical
diagonalizations cannot be converged, suggesting that the system is in the
continuum, similar to the single-atom Rabi-Stark model.

To explore the critical exponents of the SRPT, we study the finite-size
scaling law for a numerical analysis of the scaling behavior. The general
finite-size scaling ansatz for an observable $Q$ around the critical value
is given as 
\begin{equation}
Q=|1-\frac{g}{g_{c}}|^{\beta_{Q}}F(|1-\frac{g}{g_{c}}|N^{\frac{1}{\nu}})
\label{scalEq}
\end{equation}
where $F(x)$ is a scaling function, $\nu$ is the correlation length
exponent, and $\beta_{Q}$ is a scaling exponent dependent on $Q$. At the
critical value $g_{c}$ the finite-size scaling becomes $Q\propto
N^{\gamma_{Q}}$ with $\gamma_{Q}=-\beta_{Q}/\nu$.

\begin{table}[h]
\caption{Scaling exponents using the numerical calculation for the lowest
excitation energy $\protect\epsilon$, mean photon number $N_{ph}$ and the
variance of the position quadrature $\Delta x$ around the critical value $%
g_{c0}^{A_{2}}$. }
\label{exponents}%
\begin{tabular}{p{2cm}p{1.5cm}p{1.50cm}p{1.5cm}p{1.5cm}p{-0.50cm}}
\hline\hline
exponent & $\epsilon$ & $N_{ph}$ & $\Delta x$ &  &  \\ \hline
$\beta_{Q}$ & 1/2 & 1 & -1/4 &  &  \\ 
$\gamma_{Q}$ & -1/3 & -2/3 & 1/6 &  &  \\ \hline\hline
\end{tabular}%
\end{table}

We calculate the scaling exponents of the lowest excitation energy $\epsilon$%
, mean photon number $N_{ph}=\left \langle a^{\dagger}a \right \rangle /N$
and the variance of the position quadrature $\Delta x$ of the field $%
x=a+a^{\dagger}$, which are obtained by using the numerically exact
calculation. Near the critical point, the excitation energy vanishes as $%
\epsilon\propto|g-g_{c0}^{A_{2}}|^{1/2}$, while $N_{ph}$ and $\Delta x$
diverge as $N_{ph}\propto|g-g_{c0}^{A_{2}}|$ and $\Delta
x\propto|g-g_{c0}^{A_{2}}|^{-1/4}$, respectively. Table~\ref{exponents} show
different scaling exponents $\beta_{Q}$ and $\gamma_{Q}$ of the observables $%
\epsilon$, $N_{ph}$ and $\Delta x$. However, the critical exponent $\nu=2/3$
is universal and independent on observables.

\begin{figure}[tbph]
\centerline{\includegraphics[scale=0.28]{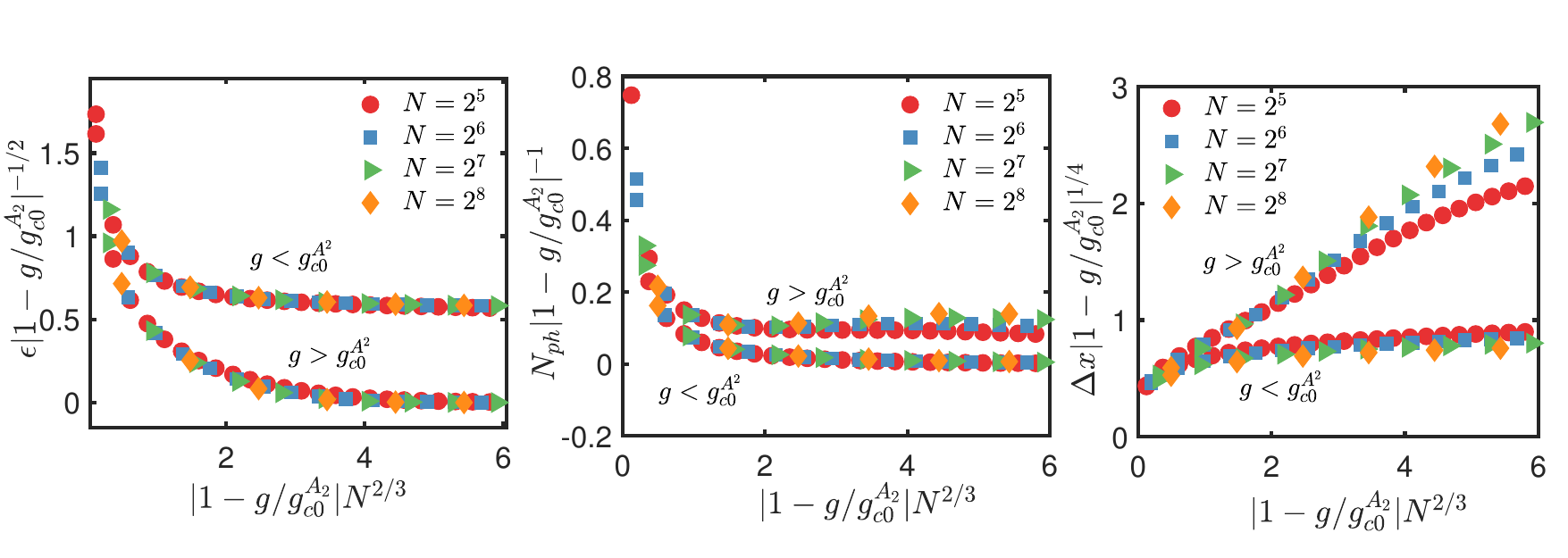}}
\caption{(Color online) Finite-size scaling of $\protect\epsilon,N_{ph}$ and 
$\Delta x$ for different size $N=2^{5},2^{6},2^{7},2^{8}$ for anisotropic
DSM with $A^{2}$ term. The parameter $\Delta/\protect\omega=0.5,U/\protect%
\omega=0.5,\protect\tau=2.5$ and $\protect\kappa=1.2$.}
\label{scalingA2}
\end{figure}

As expected, the scaling function $F(x)$ in Eq.~(\ref{scalEq}) is universal
with the critical exponent $\nu=3/2$. The finite-size scaling function of
the exictation energy can be given as $%
\epsilon|g-g_{c0}^{A_{2}}|^{-1/2}=F(|g-g_{c0}^{A_{2}}|N^{2/3})$. Fig.~\ref{scalingA2}(a) shows that all the curves of different size $N$ collapse into
a single curve in the critical regime. An excellent collapse of the scaling
function of $N_{ph}$ and $\Delta x$ are also observed in Fig.~\ref{scalingA2}(b) and (c), respectively. The numerical results confirm the validity of the
universal exponents of the anisotropic Dicke model, which belongs to the
same universality class of the conventional Dicke model.

\section{Finite-temperature phase transitions}

We investigate the superradiant phase transition at a finite temperature,
which is crucial to understand the experiment findings and
temperature-dependent physics.

\begin{figure}[tbph]
\centerline{\includegraphics[scale=0.42]{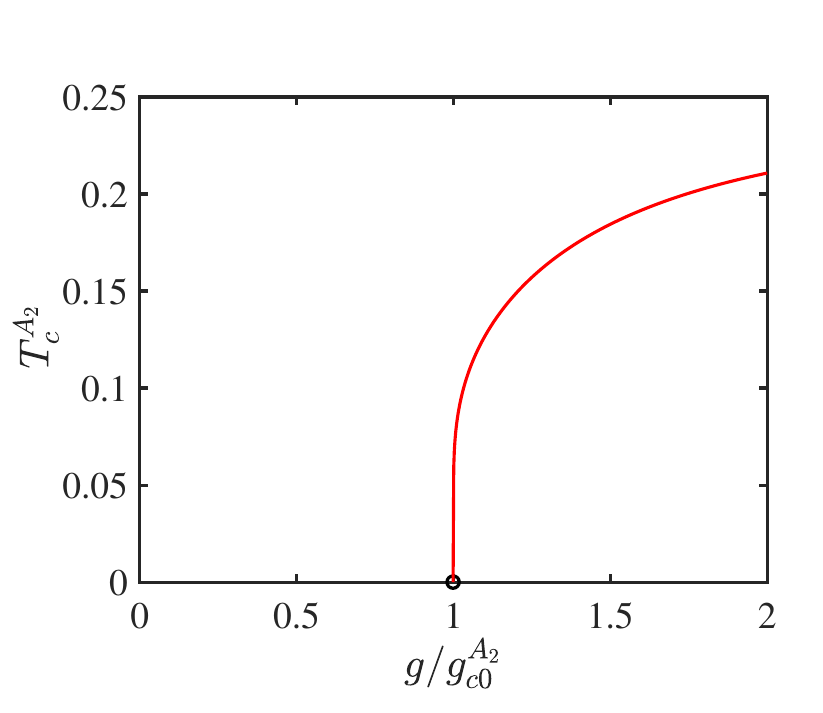}}
\caption{The critical temperature line of $T_{c}$ for the finite-temperature
phase transition at $T>0$, terminating at the quantum critical point $%
g=g_{c0}^{A_{2}}$. The parameters are $\Delta=0.5/\protect\omega,U=0.5,%
\protect\tau=1.5$ and $\protect\kappa=1.2$.}
\label{A2ani2DFig}
\end{figure}
For a finite temperature $T$, the partition function is $Z=\text{Tr}%
[e^{-\beta H}]=\text{Tr}[e^{-\beta h_{i}}]^{N}$ for the Hamiltonian in Eq.~(\ref{H2}) with $\beta=1/T$, where $H=\sum_{i}^{N}h_{i}$ is written in terms
of the Hamiltonian of each atom coupled to the cavity $h_{i}=\frac{\Delta}{2}
\sigma_{i}^{z}+\frac{U}{2N}a^{\dagger}a\sigma_{i}^{z}+\frac{g}{\sqrt{N}}%
(a^{\dagger}\sigma_{-}^{i}+a\sigma_{+}^{i})+\frac{g\tau}{\sqrt{N}}(a\sigma_{+}^{i}+a^{\dagger}\sigma_{-}^{i})$, where $\sigma_{z,\pm}^{i}$ are Pauli matrices for each atom. Using the mean-field
approximation $a\rightarrow\alpha$, the Hamiltonian $h_{i}$ can be given as 
\begin{equation}
h_i(\alpha)=\frac{\Delta}{2}\sigma_{i}^{z}+\frac{U}{2N}\alpha^{2}\sigma_{i}^{z}+\frac{g%
}{\sqrt{N}}\alpha(\sigma_{-}^{i}+\sigma_{+}^{i})+\frac{g\tau}{\sqrt{N}}%
\alpha(\sigma_{+}^{i}+\sigma_{-}^{i})
\end{equation}
The reduced free energy is defined as $F=-\ln Z/(N\beta)$. With the mean
field partition function $Z(\alpha)$, the free energy is given by 
\begin{equation}
F(\alpha)=(\omega+4D)\alpha^{2}-\frac{N}{\beta}\ln(2\cosh\beta E)
\label{FA2}
\end{equation}
where $E=\sqrt{4N\alpha^{2}g^{^{\prime }2}+\left(N\Delta
+U\alpha^{2}\right)^{2}}/2N$ is eigenvalue of $h_{i}(\alpha)$. The critical
value is obtained by the condition $\partial ^2
F/\partial\alpha^2|_{\alpha=0}=0$ 
\begin{equation}
g_{c}^{A_{2}}=\sqrt{\frac{\Delta\omega-\frac{1}{2}\Delta U\tanh\left(\frac{%
\beta\Delta}{2}\right)}{(\tau+1)^{2}\tanh\left(\frac{\beta\Delta}{2}%
\right)-4\kappa}}.  \label{gcA2}
\end{equation}
In particular, $g_{c}^{A_{2}}$ is valid only if $\Delta\omega-\frac{\Delta U}{2}\tanh\left(\frac{\beta\Delta}{2}\right)\geq0$ and $%
(\tau+1)^{2}\tanh\left(\frac{\beta\Delta}{2}\right)-4\kappa>0$. The critical
value depends on the temperature $T$, and reduces to the critical value at
zero temperature in Eq.~(\ref{gc_ADS}).

For finite-temperature SRPT, the critical temperature for SRPT is obtained
as 
\begin{equation}
T_{c}^{A_{2}}=\frac{\Delta }{2\tanh ^{-1}\left[ \frac{2\Delta \omega
+8g^{2}\kappa }{2g^{2}(\tau +1)^{2}+\Delta U}\right] }  \label{TcA2}
\end{equation}%
Fig.~\ref{A2ani2DFig}  shows the phase diagram at finite temperature. The
line of $T>0$ second order phase transition terminates at the $T=0$ quantum
critical point at $g=g_{c0}^{A_{2}}$.

\begin{figure}[tbph]
\centerline{\includegraphics[scale=0.3]{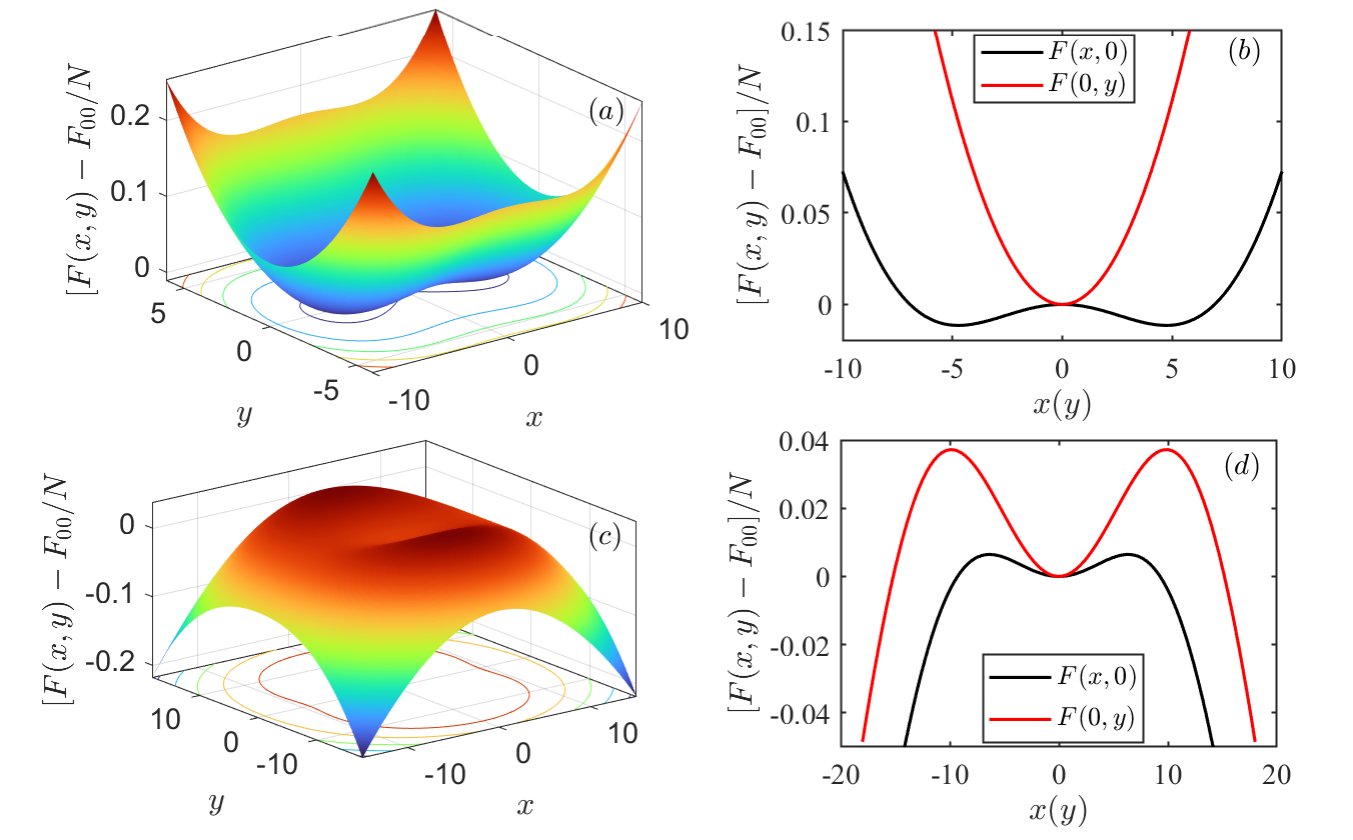}}
\caption{The free energy per atom $[F(x,y)-F_{00}]/N$ as a function of
complex $\protect\alpha =x+iy$ for $N=10^{3}$ atoms with $T/\protect\omega %
=0.5,\Delta /\protect\omega =0.5,\protect\kappa =1$. Parameters in the upper
panel are $\protect\tau =2.5,U/\protect\omega =0.5$ with $%
g_{c}^{A_{2}}=0.5160$ in the low panel are $\protect\tau =1,U/\protect\omega %
=5.5$ with $g_{c}^{A_{2}}=0.2509$.}
\label{freeEfig}
\end{figure}
The free energy is crucial to capture the second-order phase transition
according to the Landau theory. We generally set a complex $\alpha =x+iy$ in
numerical calculation. For the anisotropic case enough large $\tau>1$ in Fig.~\ref{freeEfig} (a) and (b) satisfying $\omega \geqslant \frac{U}{2}\tanh \left( \frac{\beta
\Delta }{2}\right) $ and $\tanh \left( \frac{\beta \Delta }{2}\right)
>4\kappa /(\tau +1)^{2}$, the free energy $F(\alpha )$ has two local minimums
at nonzero real value of $\alpha $ for $g>g_{c}^{A^{2}}$, while it changes
to a single local minimum at $\alpha =0$ below the critical point. It  also supports the above derivation that the free energy is expressed as a function of real  $\alpha $.  It
indicates the occurrence of the second-order phase transition. However, for
the isotropic case $\tau =1$ in Fig.~\ref{freeEfig} (c) and (d), there
emerge two local maximum at two complex value of $\alpha $ for $%
g>g_{c}^{A^{2}}$. It signals an unstate state above the critical point,
indicating no superradiant phase transition. Because substituting $\tau =1$
in Eq.~(\ref{gcA2}) leads negative numerator and denominator of $%
g_{c}^{A^{2}}$ , i.e., $\omega \leqslant \frac{U}{2}\tanh \left( \frac{\beta
\Delta }{2}\right) $ and $\tanh \left( \frac{\beta \Delta }{2}\right)
<4\kappa /(\tau +1)^{2}$. The numerical results confirm the constraints of
the phase transition. It demonstrates that the finite-temperature SRPT is
possible to occur in the anisotropic Dicke model with $\tau >1$.

\begin{figure}[tbph]
\centerline{\includegraphics[scale=0.30]{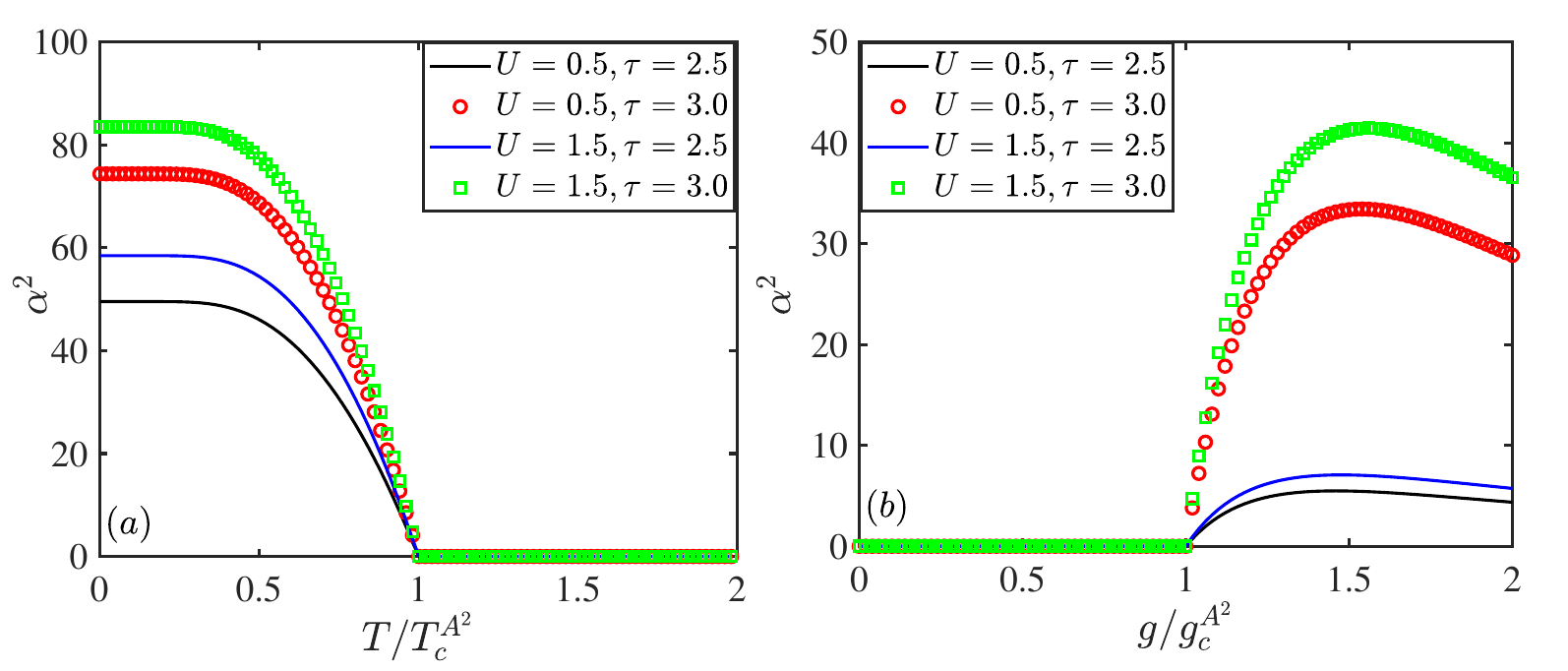}}
\caption{(a) The value of order parameter $\alpha^2$ as a function of $T/T_{c}^{A^{2}}$
at $g/\protect\omega=0.5$ for different $U$ and $\protect\tau$. (b) The
value of order parameter $\alpha^2$ as a function of $g/g_{c}^{A^{2}}$ at $T/\protect%
\omega=0.5$. The parameters are $\Delta/\protect\omega=0.5$, $\protect\kappa%
=1.2$ and $N=1000$.}
\label{FalphaA2Fig}
\end{figure}

The finite-temperature phase transition can be characterized by the
parameter $\alpha $, similar to the quantum phase transition. One
analytically obtains the value of $\alpha $ by minimization the free energy
Eq.~(\ref{FA2}) with respect to $\alpha $, which yields the equation 
\begin{equation*}
2\omega +\frac{8g^{2}\kappa }{\Delta }-\frac{\left[ 2Ng^{^{\prime
}2}+U\left( N\Delta +U\alpha ^{2}\right) \right] \tanh \left( \beta E\right) 
}{\sqrt{4N\alpha ^{2}g^{^{\prime }2}+\left( N\Delta +U\alpha ^{2}\right) ^{2}%
}}=0.
\end{equation*}%
Fig.~\ref{FalphaA2Fig} shows the order parameter $\alpha^2 $ depending on both
of the temperature $T$ and the coupling strength $g$ for different
anisotropic coupling ratio $\tau $ and the stark-shift parameter $U$. In
Fig.~\ref{FalphaA2Fig} (a), as $T$ decreases below $T_{c}^{A_{2}}$, the
system enters from the norm phase with $\alpha^2 =0$ to the superradiant
phase, and the order parameter grows from zero. For the quantum phase
transition with $T=0$, $\alpha^2 $ becomes nonzero as $g$ increase above the
critical value $g_{c}^{A^{2}}$ in Fig.~\ref{FalphaA2Fig} (b). It indicates
that there occur both of quantum and classical superradiant phase transition
in the anisotropic Dicke-Stark model with $\mathbf{A}$-square term.

\begin{figure}[H]
\centerline{\includegraphics[scale=0.28]{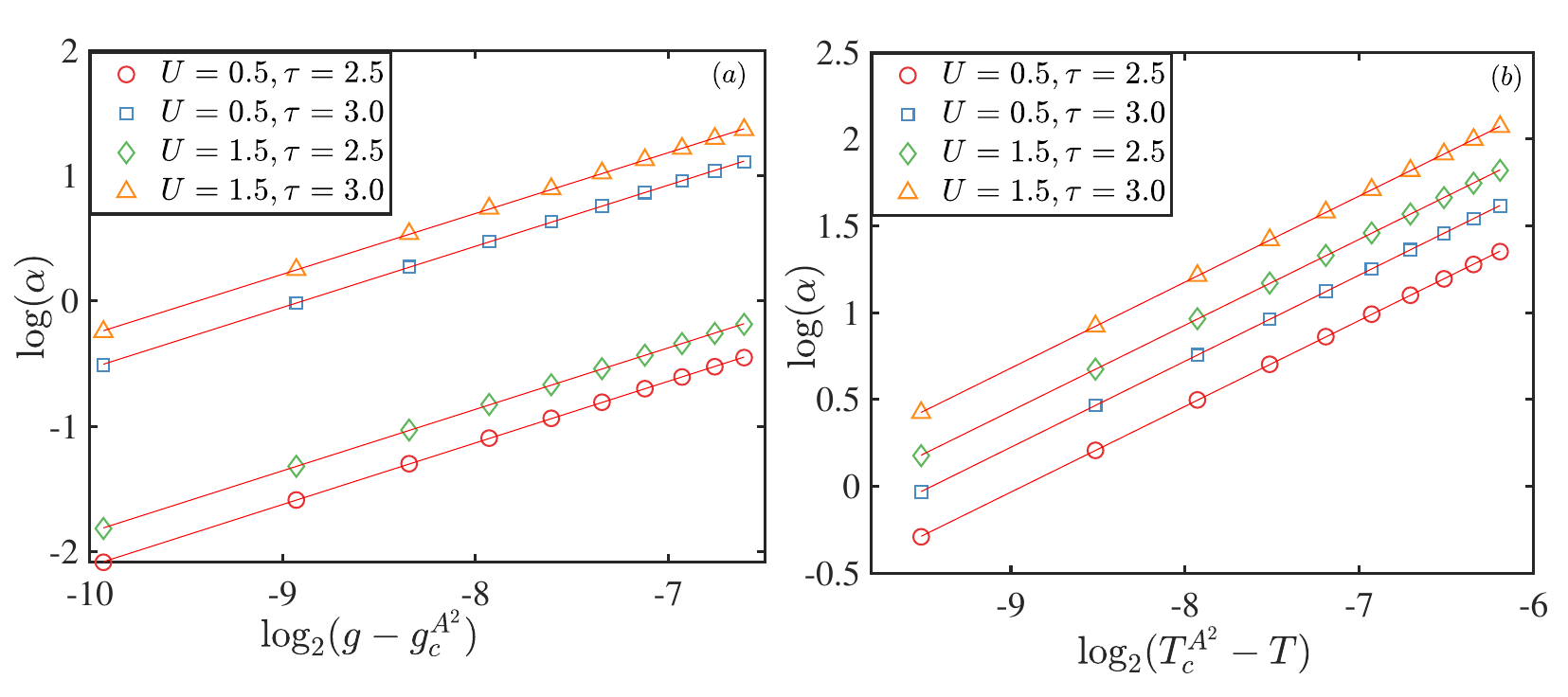}}
\caption{Scaling of $\protect\alpha$ as a function of $%
g-g_{c}^{A^{2}}$ (a) and $T-T_{c}^{A^{2}}$ (b) in log-log plot. The
parameters are $U=0.5\protect\omega,1.5\protect\omega$ , $\protect\tau%
=2.5,3.0.$, $\Delta=0.5\protect\omega,\protect\kappa=1.2$, and $N=1000$. The
slope of all fitting lines is about $\protect\beta\simeq0.5$.}
\label{loggcTc}
\end{figure}

To study the scaling exponent of the anisotropic Dicke-Stark model at finite
temperature, we expect the parameter $\alpha$ to vanish with a power-law
behavior of the form $\alpha \propto |T-T_{c}^{A^{2}}|^{\beta }$ around the
critical point. In particular for the Stark coupling $U/\omega =0$, one
easily obtains $\alpha =\sqrt{N[g^{2}-(g_{c}^{A_{2}})^{2}]f(g)}$ at zero
temperature, resulting in the scaling behavior $\alpha \propto
|g-g_{c}^{A^{2}}|^{1/2}$. In the presence of the nonlinear Stark coupling,
we numerically calculate the scaling exponent of $\alpha $ due to the
absence of the analytical expression of $\alpha $. At $T=0$, $\alpha $
scales as $\alpha \varpropto (g-g_{c}^{A_{2}})^{\beta }$ in Fig.~(\ref{loggcTc})(a), where the fitting scaling exponent is $\beta \simeq 1/2$.
Compared to the finite-temperature case in Fig.~\ref{loggcTc}) (b), one
observes $\alpha \varpropto (T-T_{c}^{A_{2}})^{\beta }$ with $\beta \simeq
1/2$. Thus, both of the quantum and classical phase transition have the same
scaling exponent, which is the same as that in the Dicke model at zero
temperature \citep{Torre2018intro}.

\section{Conclusion}

We have investigated the SRPTs at zero temperature and finite temperature
for the anisotropic DSM with an additional nonlinear stark coupling and $%
\mathbf{A-}$square term. We find that the anisotropic coupling is possible to overcome the constraints of the TRK sum rule, exhibiting the superradiant
phase transition. Moreover, the nonlinear Stark coupling reduce the critical
value, providing a experimentally feasible scheme. At finite-temperature,
the free energy for the anisotropic ratio $\tau >1$ is split into two local
minimum above the critical temperature $T_{c}$, indicating the occurrence of
the SRPT. In contrast,  the free energy of the isotropic model always has a
single local minimum  at the center, thus exclude the possibility of phase
transitions. The scaling exponent of the finite-temperature phase transition
is the same as that in the Dicke model at zero temperature. Our study opens
a window for the achievement of the superradiant phase transition in the
light-matter coupling systems.

\begin{acknowledgments}
This work is supported in part by the National Natural Science Foundation of China (Grants No. 11834005, No. 12075040, No. 12347101 and No. 2022M720387, No. 12305009). National Key R\&D Program of China 2022YFA1402701.
\end{acknowledgments}

\end{document}